# Multilayer Edge Molecular Devices Based on Plasma Oxidation of Photolithographically Defined Bottom Metal Electrode

Pawan Tyagi,

Department of Chemical and Materials Engineering, University of Kentucky, Lexington, Kentucky-40506, USA

Current Address: School of Engineering and Applied Science, University of the District of Columbia, Washington DC-20008, USA

**Abstract:** A multilayer edge molecular electronics device (MEMED), which utilize the two metal electrodes of a *metal-insulator-metal tunnel junction* as the two electrical leads to molecular channels, can overcome the long standing fabrication challenges for developing futuristic molecular devices. However, producing ultrathin insulator is the most challenging step in MEMED fabrication. A simplified molecular device approach was developed by avoiding the need of depositing a new materiel on the bottom electrode for growing ultrathin insulator. This paper discuss the approach for MEMED's insulator growth by one-step oxidation of a tantalum (Ta) bottom electrode, in the pholithographically defined region; i.e. ultrathin tantalum oxide (TaOx) insulator was grown by oxidizing bottom metal electrode itself. Organometallic molecular clusters (OMCs) were bridged across 1-3 nm TaOx along the perimeter of a tunnel junction to establish the highly efficient molecular conduction channels. OMC transformed the asymmetric transport profile of TaOx based tunnel junction into symmetric one. A TaOx based tunnel junction with top ferromagnetic (NiFe) electrode exhibited the transient current suppression by several orders. Further studies will be needed to strengthen the current suppression phenomenon, and to realize the full potential of TaOx based multilayer edge molecular spintronics devices.

**Introduction**: Molecule based electronics and spin devices are highly promising to give novel computational strategies and ultimate device miniaturization. Molecular devices have been produced by various methods [1]. The common approaches are: (a) sandwiching self-assembled molecular monolayer between two metal electrodes, (b) bridging nm gap of the metal break-junction with molecule(s), and (c) bridging the molecular conduction channels across a tunnel junction's insulator [1, 2]. The last approach is based on utilizing a prefabricated multilayer tunnel junction's edges as a test bed, and hence can be justifiably termed as multilayer edge molecular electronics device (MEMED) [2]. MEMED approach has several advantages. (a) It enables a variety of control experiments for the reliable study of molecular conduction. (b) MEMED can use a large number of metal electrodes. (c) MEMED approach utilizes simplified fabrication protocol involving conventional micro-fabrication tools, and (d) MEMED can be easily transformed into spintronics and optoelectronics devices. Several forms of MEMEDs have been developed [3-7]. The key MEMED systems are compared in figure 1 (Fig.1 a-e). All MEMEDs required the fabrication of a metal-insulator-metal tunnel junction with the exposed side edges; subsequently, molecular conduction channels are bridged across the insulator to create preferred conduction pathways. A critical review of MEMED approaches has been published elsewhere [2]. We have previously developed a liftoff based MEMED approach [7]. Under this approach a liftoff step was utilized to create exposed sides of a tunnel junction [7, 8]. Liftoff based MEMED scheme [7] avoided the need of etching step to remove unwanted insulator; etching step was integral part of previously developed methodologies [3-5]. *But, the growth of insulator is still quite challenging for MEMED fabrication* [2]. In the previous cases, it was critical to deposit the right thickness of high quality insulator on the bottom electrode. To keep the leakage current low several groups deposited 3-7 nm thick insulators [2]. However, utilization of the thicker insulator only permitted the use of long molecular channels, which could bridge across the 3-7 nm thick insulators [2]. To date such long molecular channels only increased the charge transport of a prefabricated tunnel junction; no useful switching mechanism was shown with MEMEDs. For the development of MEMED based logic and memory devices, one promising route is to utilize 1-3 nm long molecules with greater functionality. To do so it is critical to reduce the inter-electrode gap in 1-3 nm range. The liftoff-based MEMED utilized a 2 nm thick alumina (AlOx) insulator [2]. One main drawback of this approach is that the unpredictable notches along the AlOx insulator can increase the effective insulator thickness. Furthermore, it is utmost important to control the aluminum thickness, and photolithography parameters.

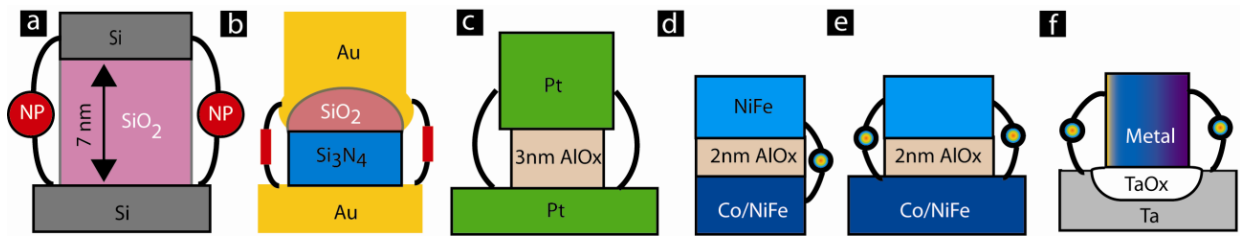

Fig.1: Summary of MEMED designs: (a) Si/SiO$_2$/Si junction's exposed edges were produced by chemical etching; hybrid molecular conduction channels were composed of gold nano particle (NP) and tethering molecules. (b) Tunnel junction with bilayer insulators was chemically etched to produce exposed side edges; insulating gap was bridged by concatenating three molecules. (c) Tunnel junction with a 3 nm AlOx was chemically etched to produce exposed side edges; gap between the new electrodes was bridged by the single molecular channel. (d) Magnetic tunnel junctions were ion milled to create exposed side edges; organometallic molecular clusters (OMCs) were bridged across the inter-electrode gap. (e) Exposed side edges were produced by liftoff step in a liftoff-MEMED with AlOx insulator; the inter-electrode gap was bridged by the OMCs. (f) Simplified liftoff-MEMED approach involved direct oxidation of the bottom electrode to avoid the deposition of additional material for insulator growth. TaOx insulating gap was bridged by the OMCs.

Here, a simplified MEMED approach is demonstrated (Fig. 1f). MEMEDs were produced by modifying the insulator growth scheme. Under the new approach, unlike other MEMEDs [3-5] no additional insulator or Al like metal for insulator growth was deposited on to the bottom electrode. Therefore, there is no need to etch extra area of insulator to produce a MEMED. The etching of extra insulator is one of the most challenging steps in MEMED fabrication approach, and may adversely affects the stability of ultrathin insulator or metal electrodes. Current scheme also overcomes one key limitation of our previously developed liftoff based MEMED fabrication approach. In the liftoff based MEMED approach we observed the side notches along the edges of AlOx insulator; these side notches disallowed molecular channels to bridge across the insulator in the affected area, hence reducing device yield.

To grow insulator for the present MEMED (Fig. 1f) a tantalum (Ta) bottom metal electrode was oxidized in a well-defined area. Pholithography was used to define the *region to be oxidized*. Subsequently, top metal electrode was deposited right over the oxidized region to produce Ta/TaOx/Metal tunnel junction with exposed sides. As an important distinction, TaOx based MEMED possessed inter-electrode gap in the plane of the substrate (Fig. 1f); however, in other MEMED approaches inter-electrode gap was perpendicular to the substrate's plane (Fig. 1a-e) [3-5]. Utilization of a top ferromagnetic electrode in a TaOx based MEMED can easily enable the fabrication of a multilayer edge molecular spintronics device (MEMSD). This chapter elaborates the fabrication details and the electrical characterization of a TaOx-based MEMED.

**Experimental details**:

The Ta/TaOx/Metal electrode fabrication scheme is summarized in figure 2. A (100) silicon (Si) wafer with 100 nm thick thermally grown silicon di-oxide was used as a substrate. These oxidized Si samples were cleaned with acetone, isopropyl alcohol and de-ionized water, respectively; after solvent cleaning samples were dried in nitrogen jet. Photolithography step defined the bottom electrode geometry. For photolithography, Shipley 1813 photoresist was spin coated at 3000 rpm rotation on substrate. Next, soft baking was performed for 1-3 min at 90-100 ºC. UV exposures with Karlsuss mask aligner through a photomask produced the required pattern. Subsequently, photoresist was developed in MF-319 alkaline developer. A 10-12 nm thick Tantalum (Ta) bottom electrode was deposited (Fig. 2a), in AJA International sputtering machine. In sputtering chamber, $2\times10^{-7}$ torr base pressure was maintained before Ta deposition. For Ta deposition, 100 W was supplied to the DC sputtering gun. During deposition, 1 mtorr argon pressure was maintained using variable aperture throttle valve. Subsequently, liftoff was conducted in Shipley resist remover to produce a patterned bottom electrode.

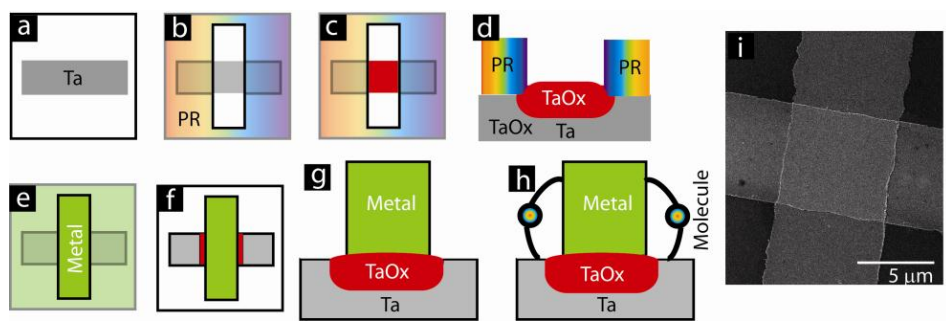

Fig. 2: Fabrication of TaOx based MEMED: (a) Deposition of bottom Ta electrode, (b) photolithography to define region for the plasma oxidation of bottom electrode and the deposition of top electrode. (c) Plasma oxidation of bottom Ta electrode (d) cross sectional view of Ta with TaOx, (e) deposition of top metal film, (f) liftoff produced tunnel junction with exposed sides. Cross-sectional schematic view of TaOx based MEMED (g) before and (h) after bridging OMCs across TaOx. (i) SEM image of a typical TaOx based junction.

In the second photolithography step a trench in photoresist was created *to only expose the bottom electrode in the pre-defined area*, (Fig. 2b). The second photolithography step was conducted using the same parameters as used for the first photolithography step for the deposition of Ta bottom electrode. However, soft backing was performed at 100-110 ºC, depending upon the photoresist age and performance. Plasma oxidation of the Ta electrode through a photoresist window was performed (Fig. 2c), within the sputtering chamber. Plasma oxidation for 30-90 seconds duration was carried out at 60 mtorr pressure of argon and oxygen mixture (Argon:Oxygen::1:1) and 20 W RF substrate bias [7].

Oxygen plasma burns organic matters. Hence, it is expected that the burning of photoresist protection, which is a hydrocarbon, during plasma oxidation compete with the lateral Ta oxidation under photoresist (PR) (Fig. 2d). Additionally, according to plasma oxidation kinetics the growth rate of TaOx is the result of competition between growth of oxidized Ta and the etching of Ta due to sputtering by the bombarding ions from plasma [9]. Plasma oxidation condition was carefully optimized to have the desired thickness of the high quality TaOx insulator and to burn the photoresist protection minimally. According to AFM study of the oxidized patterns, the PR etch rate was found to be 30±4 nm per minute.

After the plasma oxidation step the top metal electrode was sputter deposited, through the same PR window as utilized for the oxidation step (Fig. 2e). It is noteworthy that deposition of the top electrode only covered the exposed region of the TaOx, not that region which is under the PR protection (Fig. 2f). This reason prevented the short circuit between top and bottom electrodes. However, the effective gap between two electrodes was evidently smaller than the physical length of molecular clusters [10]. Various metals, such as Ta, palladium (Pd) and NiFe (with Ni:Fe: 80::20 composition), were utilized as a top electrode. All these metal electrodes showed excellent stability against all the chemicals used during MEMED fabrication. NiFe especially, not only showed stability against the electrochemical molecular self-assembly step but also was stable in the ambient conditions [11]. XPS studies revealed that after oxidation NiFe surface only had oxidized iron atoms, while Ni remained in elemental state [11]. Liftoff step produced a tunnel junction with exposed sides, where uncovered TaOx dictated the inter-electrode distance (Fig. 1f). It is noteworthy that after liftoff (Fig. 2 f-g) we observed a nonlinear tunneling electron transport on bare tunnel junction, ensuring the disconnection between two metal electrodes.

In order to transform a TaOx based tunnel junction into a molecular device, molecular conduction channels were bridged along the exposed edges (Fig. 2 h). Figure 2i shows the scanning electron micrograph of a typical Ta/TaOx/Ta metal junction. The organometallic molecular clusters (OMCs) was used for the creation of molecular channels [10]. These OMCs exhibited S=6 spin state in bulk form at <10 K, and possessed cyanide-bridged octametallic molecular clusters, $[(pzTp)Fe^{III}(CN)_3]_4[Ni^{II}(L)]_4[O_3SCF_3]_4$ [(pzTp) = tetra(pyrazol-1-yl)borate; L = 1-S(acetyl)tris(pyrazolyl)decane] chemical structure [10]. Via thiol functional group, an array of molecular clusters was covalently-linked onto the Ta and top metal electrodes [12]. For the molecule attachment, TaOx based tunnel junction samples were immersed in a dichloromethane solution of 0.1 mM molecular concentration. An alternating ±100 mV bias with time interval of 0.002 seconds for 2 min was applied between the two metal electrodes [7]. Next, molecular devices were rinsed with dichloromethane, 2-propanol, and DI water, and then dried under a nitrogen gas stream. Transport studies of the TaOx based MEMED were performed with a Keitlhley 2430 1kW Pulse Source meter and Keithley 6430 Sub-femtoamp Source meter. During transport studies, samples were mounted on a metallic chuck in a faraday cage. Biaxial and triaxial cables electrically connected the probe needles, mounted in micromanipulators, to source meters. As a standard procedure current-voltage (I-V) measurements were performed before and after molecule attachment. I-V studies were performed in ±100 mV bias range; low bias range minimized the instability caused by high bias application. Typical TaOx tunnel junction breakdown voltage was found to be ~1.7 V.

**Results and discussion:**

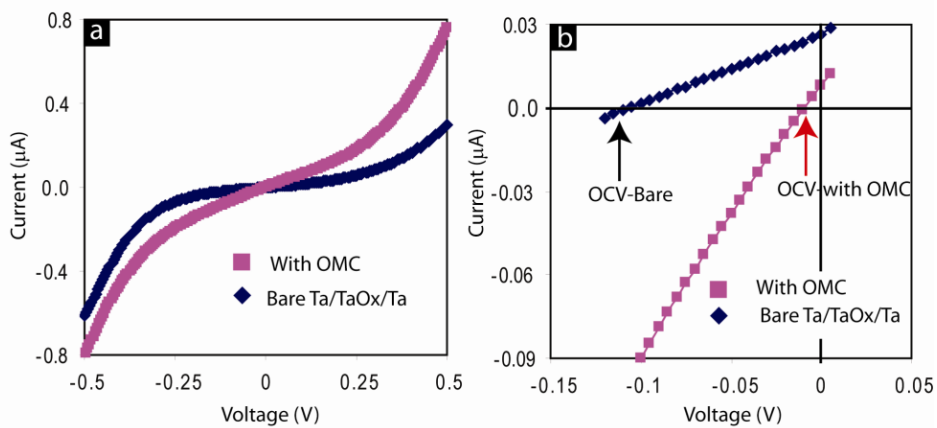

Fig. 3: OMC on Ta/TaOx/Ta tunnel junction (a) increased the overall device current and made it symmetric, and (b) reduced the photovoltaic response.

In the first attempt Ta/TaOx/Ta tunnel junction were subjected to OMC attachment. An OMC conduction bridge clearly enhanced the overall charge transport (Fig. 3a). A plasma oxidized Ta or TaOx tunnel barrier resulted in an asymmetric I-V [13]. Asymmetric transport occurs due to the inhomogeneous atomic defect profile within TaOx tunnel barrier [13]. OMC bridges made overall I-V profile to be symmetric (Fig. 3a). Molecular channels also transformed the photo response of a bare tunnel junction. A bare Ta/TaOx/Ta junction showed photovoltaic effect [13]. OMCs clearly diminished the open circuit voltage (OCV) from ~0.12 V to ~0.05V (Fig.3b). It is noteworthy that Ta/TaOx/Ta tunnel junction has a configuration of metal/insulator/semiconductor (MIS) type solar cell [14]. Here, surface of TaOx is insulating while subsurface layers are rich in Ta$^+$ type defects. Subsurface TaOx is akin to n type semiconductor. Effectively, bare TaOx tunnel junction possessed MIS configuration. After transforming a bare TaOx tunnel junction into MEMSD, OMCs dominated the conduction and photovoltaic responses associated with MIS configuration of the bare tunnel junction (Fig. 3).

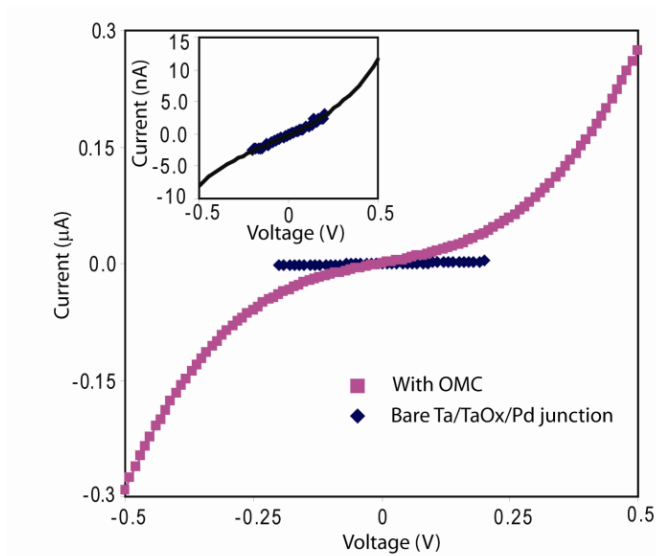

Fig. 4: Molecular clusters increased the charge transport of Ta/TaOx/Pd tunnel junction and made overall transport symmetric. In inset, the 3$^{rd}$ order polynomial fit on low bias current data of bare junction shows that before molecule attachment I-V was asymmetric.

The TaOx based MEMED enable the use of arbitrarily chosen top metal electrode. The TaOx based MEMED were produced with palladium (Pd) as the top metal electrode (Fig. 4). Here, molecular clusters increased the over all transport. Moreover, bridging of molecular clusters made I-V profile symmetric. To avoid high bias induced instability, transport data before molecule attachment was frequently recorded for the low bias range only [15]. Afterwards, the extrapolation of the low bias data with a 3$^{rd}$ order polynomial fit was used to obtain the high bias characteristic (inset of Fig. 4).

The analysis of charge transport through the TaOx based MEMED is not straight forward. The transport through TaOx tunnel junction was found to be thermally activated, before and after the bridging of molecular channels. In the present case, the application of pure tunneling model is questionable [16]. On the other hand, due to temperature induced instability we could not do temperature dependent charge transport on the same tunnel junction before and after the creation of molecular conduction channels. According to an estimate based on geometrical dimensions, a tunnel junction with 5 μm x 5μm junction area accommodated ~10,000 molecular conduction channels.

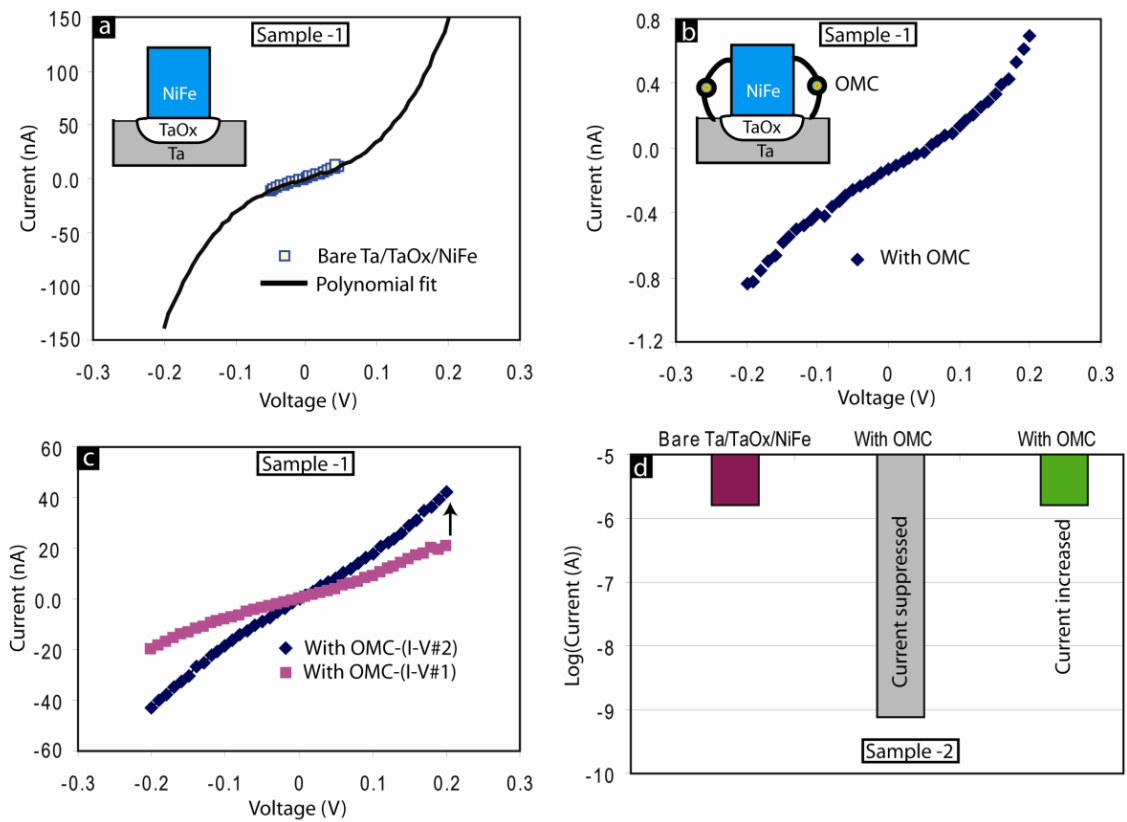

Fig. 5: Observation of transient current suppression on Ta/TaOx/NiFe: Charge transport of (a) bare tunnel junction and extrapolation of low bias experimental data using a 3$^{rd}$ order polynomial fit, (b) tunnel junction with molecular clusters in suppressed charge state, (c) suppressed charge state kept moving to higher current state with repeating I-V measurements, and (d) summary of transient current suppression on a different sample with Ta/TaOx/Pd configuration.

To produce TaOx based molecular spintronics devices NiFe ferromagnetic electrode was deposited as a top metal electrode. OMCs when bridged across TaOx of the Ta/TaOx/NiFe tunnel junction produced a current suppression by several orders (Fig. 5a-b). Noise free low current states were observed in multiple I-V measurements (Fig. 5b). However, suppressed current state gradually or sometime suddenly changed to high current state (Fig. 5c) after several I-V studies. High current state represented the effect of additional molecular conduction channels (Fig. 3-4). In several cases, the higher current level was close to the leakage current of tunnel junction, before the molecule attachment (Fig. 5d); in these cases bare tunnel junction were in general quite leaky. It is noteworthy that Ta and NiFe are highly stable against the potential chemical damage from electrochemical molecular assembly. Hence the current suppression event cannot be due to the material damage.

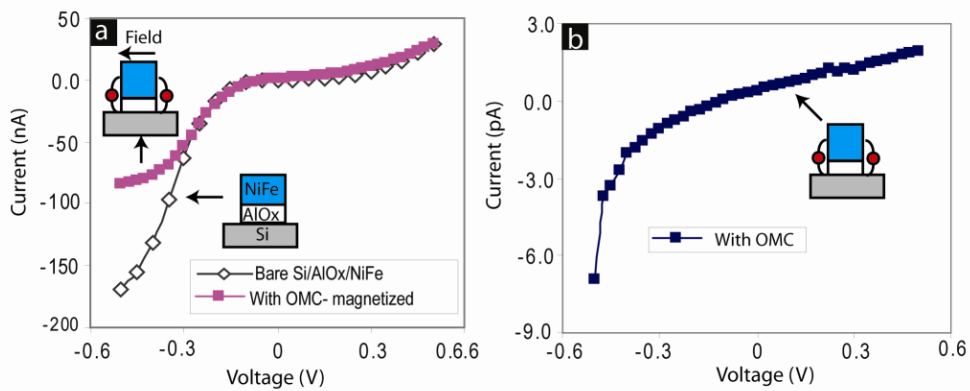

Fig. 6: Transient current suppression with Si/AlOx/NiFe: I-V measurements (a) in bare state and (b) in suppressed current state after the bridging of OMCs.

Similar transient current suppression was also observed with a different system. The Si/AlOx/NiFe tunnel junctions with OMCs (Fig. 6a-b) exhibited current suppression by ~ 4 orders (Fig. 6b). These junctions were studied under microscope to check the possibility of mechanical damage. Surprisingly, after multiple charge transport studies or in plane magnetization high current state returned; however, I-V profile was significantly different from that of a bare tunnel junction (Fig. 6a). In this case silicon and NiFe both were resistant to the chemical damages.

The underlying mechanism behind OMC induced transient current suppression is not clear yet. However, it is certainly due to the interaction between OMCs and NiFe ferromagnetic electrode, *which was present as one of the tunnel junction electrode*. In the bulk form OMC was found to possess S=6 spin state [10] at <10 K; with the increase in temperature total spin decreased. Besides this, an OMC like magnetic molecule is expected to experience significant change when interacting with a metal surface [17]. A significant changes in the spin state and rearrangement of energy level is highly likely when OMC self-assembled on a metallic surface [18]. To understand the effect of OMCs on the magnetic properties of Ta/TaOx/NiFe the magnetization studies were performed on the analogous system. For the magnetization study a Quantum design MPMS Squid magnetometer was utilized. For this study, a sample with 21,000

Pd/AlOx/NiFe tunnel junction dots was studied at 150 K. Here, Pd is a nonmagnet, hence Pd/AlOx/NiFe is akin to Si/AlOx/NiFe and Ta/TaOx/NiFe, the two configurations on which transient current suppression was observed. Surprisingly, OMC decreased the magnetization of bare Pd/AlOx/NiFe (Fig. 7a). It is noteworthy that the OMCs did not affect the magnetization of unpatterned NiFe film. OMCs only affected the magnetization of NiFe when it was the part of a tunnel junction. Presumably, OMCs enhanced the exchange coupling between NiFe and nonmagnetic metal. According to transport and magnetic studies coupling via OMCs is much stronger than that via AlOx tunnel barrier. Enhanced coupling via OMC is expected to induce new magnetic ordering in NiFe ferromagnet; NiFe's edge vicinity is presumably experienced the OMC induced coupling (Fig. 8).

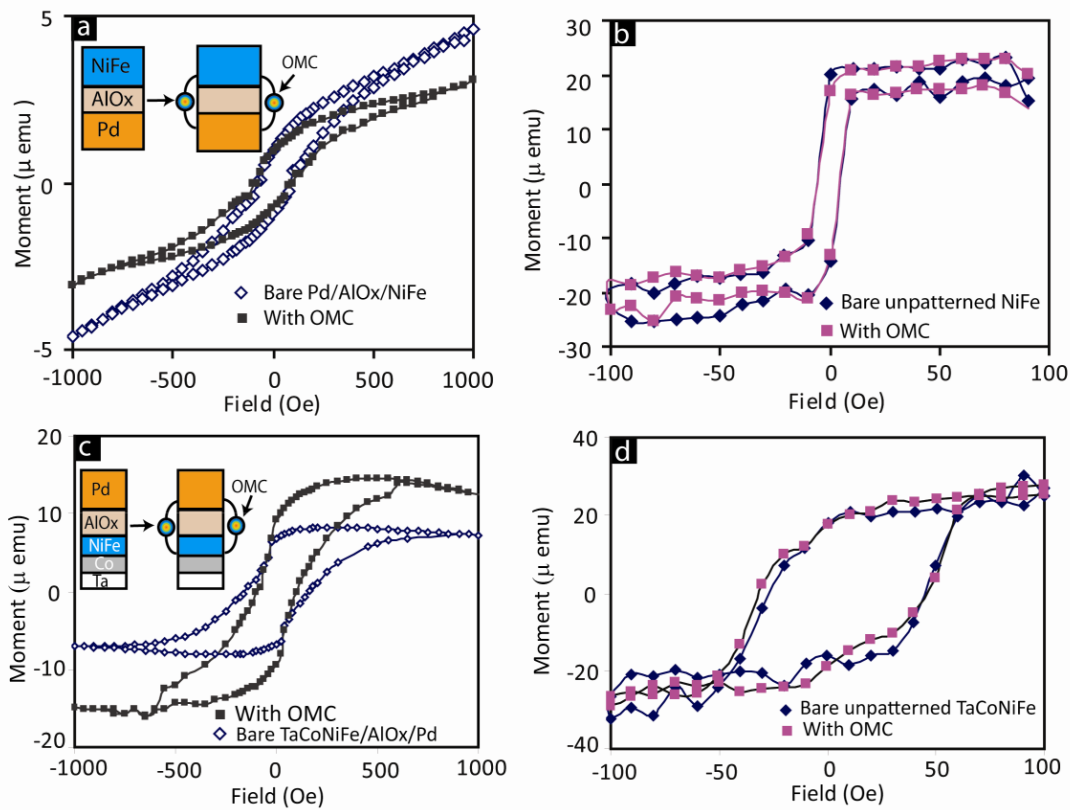

Fig. 7: Magnetization study of tunnel junctions with one ferromagnetic electrode with and without OMCs: (a) the magnetization of NiFe/AlOx/Pd reduced upon the bridging of OMCs. (b) the magnetization of NiFe unpatterned flat film did not show change in magnetization due to same OMC. (c) Magnetization of TaCoNiFe/AlOx/Pd increased upon the bridging of OMC, as shown in inset. (d) Magnetization of unpatterned TaCoNiFe composite ferromagnet remained unaffected.

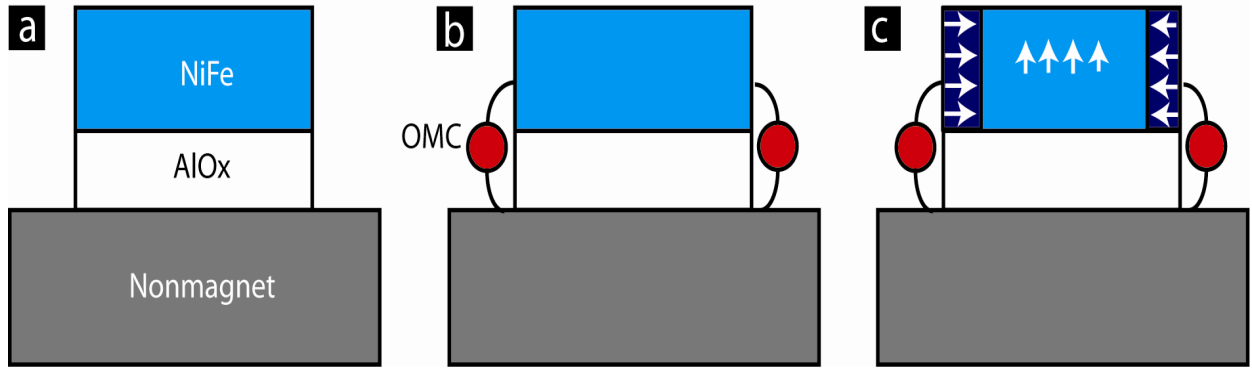

Fig. 8: Hypothesized representation of OMCs' effect on NiFe's magnetic ordering: nonmagnet-AlOx-ferromagnet in (a) bare state and (b) with OMCs. (c) Near edge regions are affected by OMCs.

Change in coupling strength between two electrodes can affect its macroscopic properties. Inter-electrode coupling strength can be enhanced in multiple ways. In present case we added OMCs channels along with AlOx insulator to enhance to enhance the inter-electrode coupling. Other approach may be to reduce the thickness of spacer between two electrodes as inter-electrode coupling strength increase exponentially with reducing thickness. In a related study, enhancing the coupling strength between two ferromagnets by reducing the nonmagnetic spacer thickness, changed the Curie temperature by > 50 K [19]. In this case ferromagnets were only few monolayers thick. In another complementary study Pasupathy et al. [20] observed that a $C_{60}$ molecular coupler between two Ni ferromagnets resulted in Kondo resonance splitting without the application of external magnetic field. It was concluded that molecule mediated coupling resulted in extremely strong, ~60 T, local magnetic field which substituted the need of applying external magnetic field to cause Kondo resonance splitting. These two studies affirmed that enhancing inter-electrode exchange coupling can yield unprecedented and surprising effects.

In the context of our observations, OMC enhanced inter-electrode exchange coupling resulted in current suppression and also remarkably influenced the magnetization response of tunnel junctions. However, we are unclear about the exact correlation between the reduction in magnetization of a tunnel junction with NiFe electrode and the current suppression on Ta/AlOx/NiFe (10 nm) and Si/AlOx/NiFe(10 nm) tunnel junctions. Study showing OMC induced magnetization reduction is of marked significance because magnetic signal recorded for this study was the average effect of ~21,000 magnetic tunnel junctions. To ascertain that magnetization result on Pd/AlOx/NiFe is not serendipitous, another system was investigated. In a related study, same OMC affected the magnetization of another magnetic layer coupled to a nonmagnetic metal via insulator. The Cobalt (Co) (3-5 nm)/NiFe (5-7 nm) bilayer magnetic electrode in Ta/Co/NiFe/AlOx/Pd exhibited an increase in magnetization when OMCs were bridged across the insulator. Here, Ta served as a seed layer, and Co was used for modifying the magnetic attributes of NiFe. It is noteworthy that OMCs bridging across AlOx, was still chemically bonded to NiFe and Pd. These magnetization studies proved that OMCs significantly changed the magnetic properties of a tunnel junction with one ferromagnetic electrode. OMCs increased the magnetization of tunnel junction with Ta/Co/NiFe electrode. However, OMCs did not affect the unpatterned Ta/Co/NiFe ferromagnetic film (Fig.7d), like the case of NiFe (Fig. 7b). It is believed that OMC mediated the intermixing of DOS of the two electrodes. This intermixing supposedly influenced the energy bands and the spin density of state of ferromagnetic electrode to influence the charge transport rate through *both,* molecule and tunnel barrier. A schematic delineating our hypothesis about the molecule induced magnetic ordering is shown in Figure 8. OMCs also produced a stable current suppression on magnetic tunnel junctions with *two magnetic electrodes*. Magnetic tunnel junction with CoNiFe/AlOx/NiFe configuration exhibited stable current suppression. The magnetization and charge transport studies on this system are discussed in chapters 4 and 5.

**Summary:** This study discussed the fabrication and characterization of TaOx based multilayer edge molecular electronic devices (MEMED). The TaOx based MEMED obviate the need of depositing separate insulator on the bottom electrode for the creation of molecular dimension inter-electrode gap. Instead, a section of bottom electrode was oxidized to produce insulating separator between two metal electrodes. The use of Ta also produced a chemical resistant bottom electrode. This approach can be used as a test bed to study a variety of molecular conduction channels. TaOx based MEMED can be produced in straightforward manner. However, a simplification in MEMED fabrication approach cost the MEMED's versatility. For instance, bottom electrode can only be made up of oxidizable metals. Oxidizable bottom electrode is required to produce insulator spacer between two metal electrodes. Moreover, oxidation of bottom electrode at the site of molecule attachment can significantly reduce the population of molecular conduction channels. Commencement of electrochemical molecule attachment step in the end, flexibility in using any metal as the second electrode, the presence of active molecules in exposed region are other major merits of our system.

Magnetic molecules were found to produce transient current suppression. Magnetization studies on the thousands of tunnel junction showed that same magnetic molecules also produced significant change in magnetic moments of bare tunnel junctions. Lack of resources, limited stability of tunnel junctions, and difficulty in maintaining consistent experimental conditions for long durations only enabled limited studies. However, a number of supporting experiments were successfully accomplished on related systems [21]. Further studies by independent researchers will be crucial to strengthen the results of this paper. It is expected that a system of magnetic tunnel junction with porphyrins or single molecular magnets conduction bridges is highly promising to observe similar or more exotic phenomenon.

**Acknowledgments:**

PT thanks Prof. Bruce J. Hinds and the Department of Chemical and Materials Engineering, University of Kentucky to enable his PhD research work presented in this manuscript. He also thanks D.F Li and S. M. Holmes for providing molecules used in this work.